\documentclass[preprint,aps,prb]{revtex4}
\usepackage{bm}
\usepackage{epsfig}
\usepackage{colordvi}
\newcommand{\nix}[1]{}

\begin{document}

%
%
%

\title{Magneto-gyrotropic photogalvanic effects due to inter-subband
absorption in quantum wells}

\author{H Diehl$^{1}$, V A Shalygin$^{2}$,   S N Danilov$^{1}$,
S A Tarasenko$^{3}$, V~V~Bel'kov$^{3}$,  D Schuh$^{1}$, W
Wegscheider$^{1}$, W Prettl$^{1}$ and S~D~Ganichev$^{1}$}

\affiliation{$^1$Terahertz Center, University of Regensburg, 93040
Regensburg, Germany}
\affiliation{$^2$St Petersburg State Polytechnic University,
195251 St Petersburg, Russia}

\affiliation{$^3$A.F.~Ioffe Physico-Technical Institute, Russian
Academy of Sciences, 194021 St.~Petersburg, Russia}



\begin{abstract}
{We report on the observation of the magneto-photogalvanic
effect (MPGE) due to inter-subband transitions in $(001)$-oriented
GaAs quantum wells. This effect is related to the gyrotropic
properties of the structures. It is shown that inter-subband absorption of
linearly polarized radiation may lead to spin-related as well as
spin independent photocurrents if an external magnetic field is
applied in the plane of the quantum well. The experimental results
are analyzed in terms of the phenomenological theory and
microscopic models of MPGE based on either asymmetric optical
excitation or asymmetric relaxation of carriers in $\bm{k}$-space. We
observed resonant photocurrents not only at oblique incidence of
radiation but also at normal incidence demonstrating that
conventionally applied selection rules for the inter-subband
optical transitions are not rigorous.}
\end{abstract}

\maketitle



Optical excitation of free carriers in a semiconductor can lead to
generation of an electric current.
In this paper we deal with the photogalvanic effects, which appear
neither due to inhomogeneity of the optical excitation nor due to
inhomogeneity of the sample (in the case of quantum well structure
we mean in-plane inhomogeneity). Such photocurrents
are caused by the
asymmetry of elementary processes of photoexcitation or carrier
relaxation resulting in a redistribution of carriers in momentum
space.
This is possible in structures of the appropriate symmetry only,
in particular, gyrotropic materials give a rich spectrum
of photogalvanic phenomena.
In gyrotropic quantum well structures (QWs) photocurrents
comprise (for a review
see~\cite{Ganichev03p935,Ivchenkobook2,GanichevPrettl}) the
linear~\cite{Gusev87} and
circular~\cite{PRL01,Bieler05,Yang06,Cho07} photogalvanic effects,
the linear and circular photon drag
effects~\cite{Luryi87p2263,Wieck90p463,Shalygin06}, the
spin-galvanic effect due to optical
excitation~\cite{Nature02,PRB03sge}, as well as  photocurrents
caused by quantum interference of one- and two-photon
excitation~\cite{Bhat00p5432,Stevens02p4382,Huebner2003,Stevens2003},
which represent  coherent photogalvanic
effects~\cite{Entin89p664,Magarill01p652}. These investigations
explored new mechanisms of photocurrent formation and gave an
experimental access to in-plane symmetry properties, spin
relaxation, and details of band structure, in particular, to
relativistic spin-orbit features of low-dimensional structures
like Rashba and Dresselhaus spin-splitting in semiconductor
quantum wells~\cite{PRL04,PRB07RD}.

An additional root of photogalvanic effects is provided by
application of an external magnetic field. The magnetic field
breaks the  time inversion symmetry  resulting in  additional
mechanisms of photocurrents. For instance, in inversion
asymmetric but non-gyrotropic  bulk crystals like GaAs,
homogeneous illumination with circularly polarized radiation does
not yield a current. However, in the presence of an external
magnetic field this effect becomes possible and has been
detected~\cite{Andrianov84p882}. Essential progress has been
achieved in the generation  of magnetic field induced
photogalvanic effects applying visible, near-, mid- and
far-infrared laser radiation (for a review
see~\cite{Ivchenkobook2,GanichevPrettl,sturman}). It has
been demonstrated that optical excitation of quantum well
structures can result in spin photocurrents, caused by the Drude
absorption due to spin-dependent asymmetric
scattering~\cite{Belkov2005,naturephysics06,PRB07MPGE} or direct
optical transitions between branches of the spin-split electron
subband~\cite{Magarill90p2064,Dmitriev91p273}, as well as in
photocurrents due to spin-independent  diamagnetic
mechanisms~\cite{gorbats,moscow,kucher,Kibis1997}.

Here we present experimental and theoretical studies of
photocurrents induced by inter-subband optical transitions in
quantum wells in the presence of an in-plane magnetic field. The
analysis gives evidence that the observed photocurrents are
related to gyrotropy of quantum wells, which is caused by bulk
and/or structure inversion asymmetry. We demonstrate that the
magneto-gyrotropic photogalvanic effect generated by direct
transitions between  size-quantized subbands is caused by a new,
so far not discussed, type of spin-independent diamagnetic
mechanism. We show that the  current  is due  to
$\bm{k}$-linear diamagnetic terms in the scattering
amplitude which describe the relaxation of photoexcited
carriers.

\section{Experimental results}

Experiments have been carried out on molecular-beam-epitaxy  grown
$(001)$-oriented $n$-doped GaAs/AlGaAs multiple (30 periods) QWs at room
 and liquid helium temperatures. Each quantum well of
$8.8$~nm, 8.2~nm, and 7.6~nm width contains a two-dimensional
electron gas with the carrier density of $3\cdot10^{11}$ cm$^{-2}$
at $4.2$~K. The quantum well widths are chosen to be around  8~nm
so that the separation  between the lowest ($e1$) and the second
($e2$) conduction subbands matches the photon energy range of a
CO$_2$-laser~\cite{PRB03sge,Hilber97p85,Tsujino00p1560}. In order
to correlate the spectral dependence of the photocurrent to the
absorption of the QWs, optical transmission measurements
have been performed  using a Fourier transform
infrared spectrometer. Direct inter-subband optical transitions in
$n$-type GaAs QWs are obtained by applying
various wavelengths of a line tunable Q-switched as well as an
transversal excited atmospheric pressure (TEA)
CO$_2$-lasers~(see, e.g.,
\cite{GanichevPrettl,Svelto}). The lasers yield linearly
polarized radiation at wavelengths
$\lambda$ between 9.2\,$\mu$m and 10.8\,$\mu$m corresponding to
photon energies $\hbar \omega$ ranging from 135~meV to 114~meV.
The Q-switched laser has  a peak power of 2~kW at a repetition
rate of 300~Hz.

The samples are irradiated at normal incidence,
i.e. along the growth direction, as well as at oblique incidence
with an angle $\theta_0$ between the light propagation direction
and the QW growth direction $z$. We  use here
Cartesian coordinates $x
\parallel [1\bar{1}0]$, $y \parallel [110]$, $z \parallel [001]$.
The external magnetic field $\bm B$ up to $1$~T is
 applied parallel to the interface plane along the $x$ or $y$ directions. In order to vary the
angle between the polarization vector of the linearly polarized
light and the magnetic field, we place a linear polarizer, made of two Brewster windows,
behind a Fresnel rhomb. After passing through the Fresnel rhomb
initially linearly polarized laser light becomes circularly polarized. Rotation of the linear
polarizer enables us to obtain linearly
polarized radiation with angle $\alpha=0^\circ \div 360^\circ$
between the vector of light polarization and the incidence plane.

\begin{figure}[!b]
\centering
\includegraphics[width=9cm]{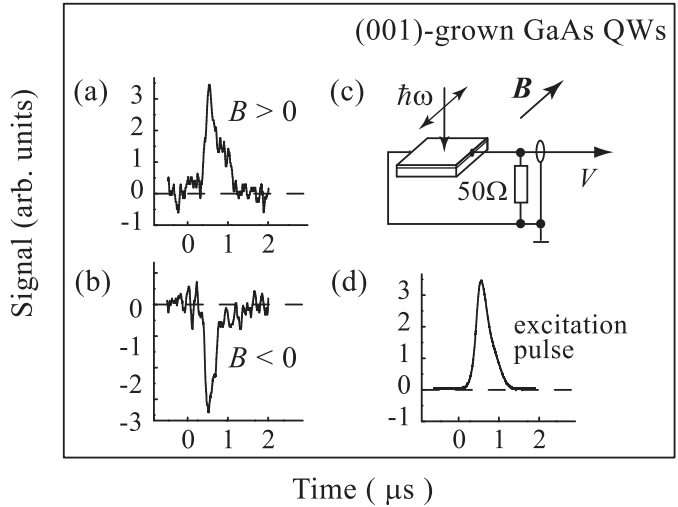}
\caption{Oscilloscope traces obtained for pulsed excitation of
(001)-grown $n$-type GaAs QWs at $\lambda = 10.27~\mu$m.
Figures (a) and (b) show magneto-gyrotropic photogalvanic
signals obtained for magnetic fields $B_y = 0.3$~T and $B_y =
-0.3$~T, respectively. For comparison in (d) a signal pulse of a
fast photon drag detector is plotted. In (c) the measurement
arrangement is sketched.}
\label{fig1}
\end{figure}

\begin{figure}[!b]
\centering
\includegraphics[width=11cm]{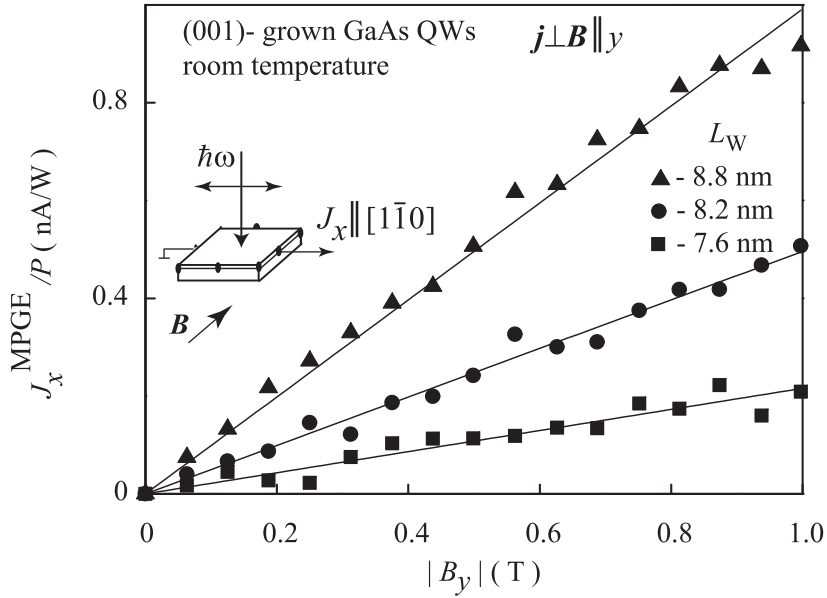}
\caption{Magnetic field dependence of the photocurrent measured in
(001)-grown GaAs QWs structures of various widths on the magnetic
field $\bm B$ parallel to the [110] axis. Optical excitation of $P
\approx 1$~kW at normal incidence is applied at three different
wavelengths corresponding to the signal maximum in each QW, see
arrows in Fig.~\protect\ref{fig3} ($\lambda_1$ for
 QW with $L_W$~=~8.8~nm, $\lambda_2$ for $L_W$~=~8.2~nm and $\lambda_3$
 for $L_W$~=~7.6~nm) . The current is measured in the
direction perpendicular to $\bm B$  and for radiation
polarized perpendicular to $\bm B$.
}
\label{fig2}
\end{figure}

To measure electric currents, two pairs of Ohmic contacts have
been centered at opposite sample edges oriented along $x$ and $y$
(see Fig.~\ref{fig1}(c)).  The photocurrent $J$ generated in the
unbiased devices is measured via the voltage drop across a
50~$\Omega$ load resistor in a closed circuit configuration (see
Fig.~\ref{fig1}(c)). The voltage in response to a laser pulse is
recorded with a fast storage oscilloscope.

Irradiation of  samples with normally incident linearly polarized
radiation in the absence of an external magnetic field causes no
photocurrent. This result agrees with the phenomenological theory
which does not allow any photocurrent at homogeneous excitation of
structures belonging to D$_{\rm 2d}$ or C$_{\rm 2v}$ point-group
symmetries relevant to (001)-grown GaAs QW
structures~\cite{Ganichev03p935,Ivchenkobook2,GanichevPrettl}. A
photocurrent response is obtained only when a magnetic field $\bm
B$ is applied. The signal is detected only in the direction
perpendicular  to the orientation of the magnetic field
independent whether  $\bm B$ is aligned along $x$ or $y$. The
basic features of the signals remain the same in both geometries,
therefore, in the following,  we give only results for
photocurrents measured in the $x$ direction and magnetic field
oriented along $y$. The signal follows the temporal structure of
the laser pulse and changes the sign upon inversion of the
magnetic field direction from $B_y
> 0$ to $B_y < 0$. Signal traces are shown in Fig.~\ref{fig1}
compared to records of a linear photon drag
detector~\cite{GanichevPrettl}. Besides the experimental geometry
with normal incidence of the radiation on the
samples, the magnetic field induced photocurrent has also been
observed at oblique incidence. We note that the excitation
of QWs at oblique incidence results in a measurable
electric current even at zero magnetic field due to
the linear photogalvanic and photon drag
effects~\cite{GanichevPrettl}. In this work we examine magnetic
field induced photocurrents $J^{MPGE}_x$ only, i.e.,
currents which reverse their sign upon switching the magnetic
field direction. In order to extract such a current contribution
from the measured total current we
determine the response to the magnetic field aligned along
the $y$ axis, $B_+$, and along
$-y$, $B_-$, and evaluate the data after
\begin{equation} \label{MPGE} J^{MPGE}_x =
\left[J_x(B_{+})-J_x(B_{-})\right]/2 \:.
\end{equation}

\begin{figure}[!b]
\centering
\includegraphics[width=11cm]{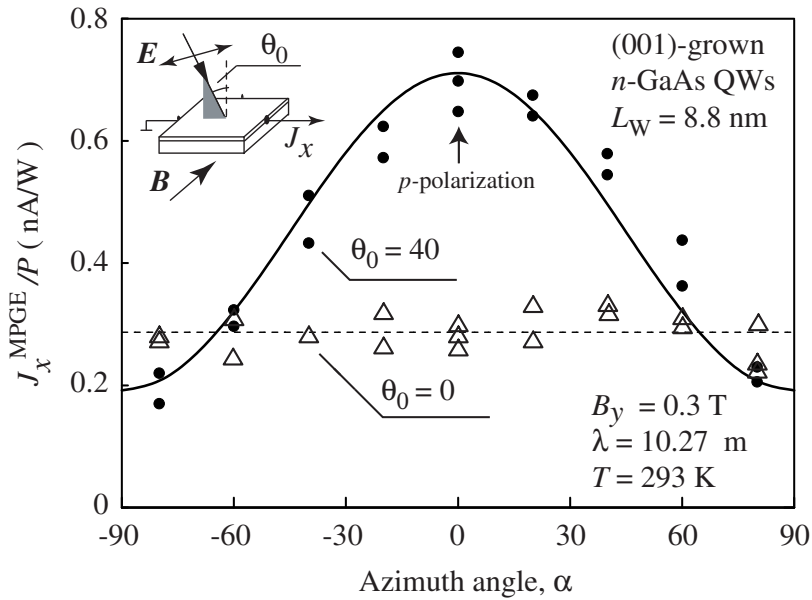}
\caption{Dependences of the magneto-induced photocurrent
$J_x^{MPGE}$ normalized by the power
$P$ on the azimuth angle $\alpha$ of the light
polarization plane measured at normal incidence
($\theta_0=0^\circ$)
 and at oblique incidence ($\theta_0=40^\circ$). In the latter case
the  plane of incidence is the ($xz$) plane. Azimuth angle $\alpha =0^\circ$
corresponds to the maximum value of  the electric field component
normal to the QW plane ($p$-polarization). Solid and dashed curves
are fits by an analytical expression
given by Eq.~(\protect \ref{K239})
taking into account the light refraction and absorption
in QW structure (see Eq.~(\protect\ref{K1})).
}
\label{fig4}
\end{figure}

Magnetic field and polarization dependencies of
$J_x^{MPGE}$ are presented in
Figs.~\ref{fig2} and ~\ref{fig4}. As is shown in Fig.~\ref{fig2},
the photocurrent exhibits linear dependence on the magnetic field
strength in all investigated samples.
Figure~\ref{fig4} demonstrates the essential difference between the
magnetic field induced
photocurrents excited at normal and at oblique incidence.
While at normal incidence the current is almost independent
 of the radiation polarization, at
oblique incidence the magneto-induced photocurrent gets
polarization-dependent: it reaches maximum for the radiation
polarized in the incidence plane ($p$-polarization, $\alpha =
0^{\circ}$) and minimum for the orthogonal polarization where
radiation electric field has no component normal to the QW plane
($s$-polarization, $\alpha = 90^{\circ}$).

\begin{figure}[!b]
\centering
\includegraphics[width=8cm]{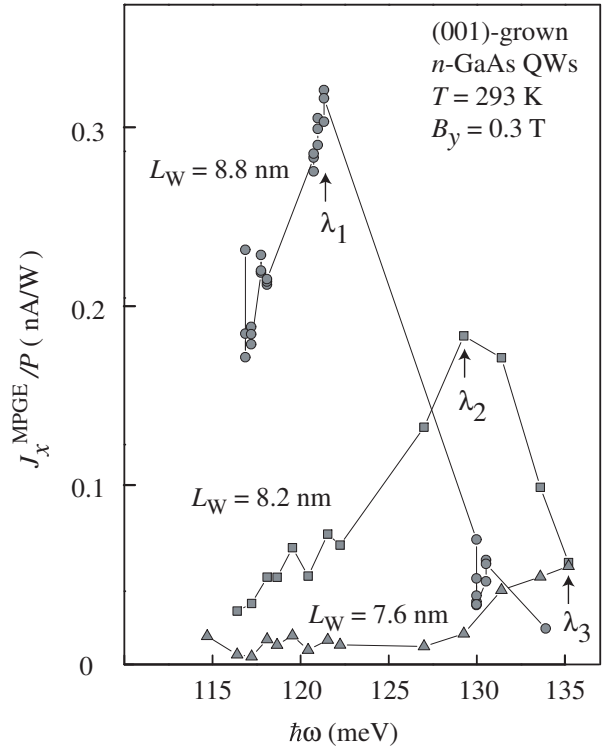}
\caption{Spectral dependences of the magneto-induced photocurrent
$J_x^{MPGE}$ normalized by the power
$P$ at the magnetic field $B_y=0.3$~T. The dependences are measured for three (001)-grown
GaAs/AlGaAs samples with various QW widths under normal
incidence of the light at room temperature. Arrows indicate
wavelengths used in Fig.~\protect\ref{fig2}.} \label{fig3a}
\end{figure}

\begin{figure}[!b]
\centering
\includegraphics[width=10cm]{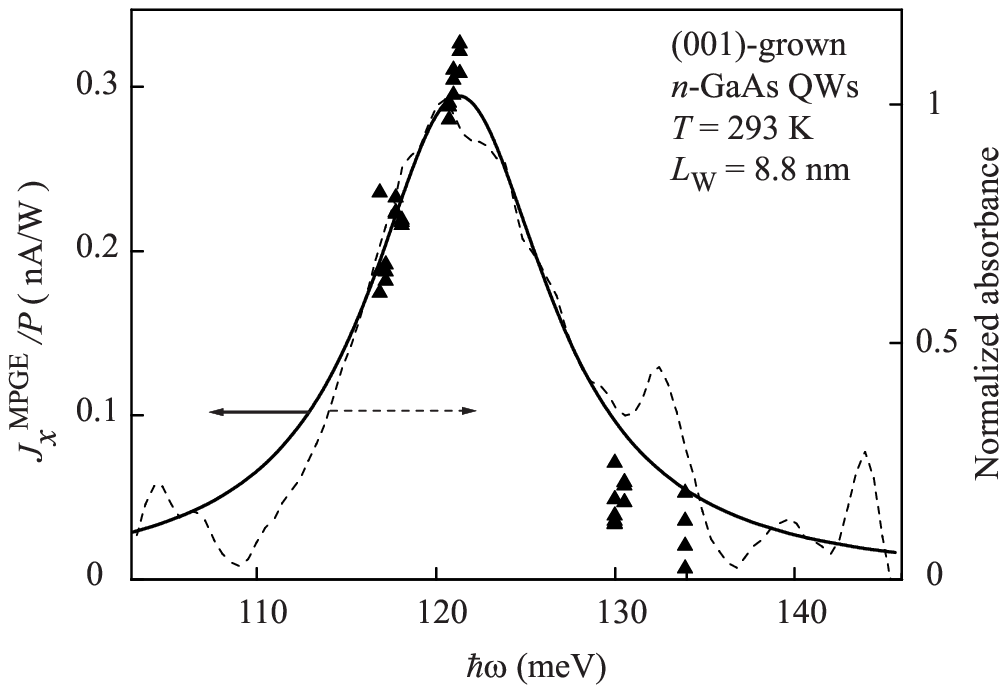}
\caption{Spectral dependence of the magneto-induced photocurrent
$J_x^{MPGE}$ normalized by the power
$P$ at the magnetic field $B_y~=~0.3$~T. The current is measured
under normal incidence of the light at room temperature. Solid
curve is fit to experimental data (triangles) by a Lorentz
function. The
dashed curve shows the normalized absorbance spectrum measured in
the multiple pass geometry.} \label{fig3}
\end{figure}

Figure~\ref{fig3a} shows the spectral dependences of the
photocurrent obtained in the  range of photon energies accessible
with the CO$_2$-lasers. The data are obtained at normal incidence
for  a constant magnetic field $B = 0.3$~T in QW structures of
various QW widths $L_W$. It is seen that the photocurrent has a
resonant character and the peak of the resonance shifts to the
higher energies for narrower QWs. In Fig.~\ref{fig3} the observed
current for a sample with QWs of 8.8~nm width is
plotted as a function of photon energy $\hbar \omega$ together
with the absorption spectrum.
The spectral dependence of the photocurrent corresponds to that of
inter-subband absorption measured by means of the Fourier
transform infrared spectroscopy  and can be well fitted by  a
Lorentzian. The photocurrent resonance shifts to shorter
wavelength (blue shift) and narrows if the temperature is reduced
from room to liquid helium temperature. All the
observed features, the coincidence of the photocurrent and
the absorbance spectra, the shift of the spectral
position of the photocurrent resonance by variation of the QW
width,
as well its temperature behaviour
demonstrate that the observed photocurrent is caused by the
inter-subband transitions.

Now we analyze the dependence of the MPGE photocurrent on the angle
of incidence $\theta_0$ and the
azimuth angle $\alpha$
considering that the QW structure represents a absorbing uniaxial
 medium with the optical axis perpendicular to the
structure surface. We assume that in the investigated
samples the anisotropy of the refractive index is
sufficiently small ($n_z\approx n_{\parallel}\approx n$), but the
anisotropy of the absorbance  is strong ($\eta_z \gg
\eta_{\parallel}$), where the subscripts  $z$ and $\parallel$
correspond to the radiation polarized along the growth direction
$z$  and parallel to the  QW plane, respectively. In the case of
linearly polarized radiation, the polarization dependence of the
structure absorbance $\eta(\alpha,\theta)$ following to
Ref.~\cite{BornMWolfE1999} and Fresnel laws is described by
\begin{equation}
\eta(\alpha,\theta )= t_p^2 \cos^2\alpha (\eta_{\parallel}
\cos^2\theta + \eta_{z} \sin^2\theta) + t_s^2 \eta_{\parallel}
\sin^2 \alpha \:,
 \label{K1}
\end{equation}
where $\theta$ is the refraction angle in the QW structure,
$\sin{\theta}=\sin{\theta_0}/n$, $t_p$ and $t_s$ are the
transmission coefficients through the sample surface for  $p$- and
$s$-polarized components of the light electric field,
respectively~\cite{Guenther1990}.

We suppose that the polarization dependence of the MPGE
current density
is determined solely by the
polarization dependence of the radiation absorbance and,
therefore, has the form
\begin{equation}
j^{MPGE}_x = \gamma B_y \frac{nc}{4\pi}E^{2}_{0}
\eta(\alpha,\theta ),
 \label{K239}
\end{equation}
where $\gamma$ is a parameter, $E_{0}$ is the electric field
amplitude of incident light and $c$ is the light velocity.
While in the experiments the electric current
$J^{MPGE}_x$is measured, in the theoretical consideration the
current density $j^{MPGE}_x$ is used which is proportional to the
current $J^{MPGE}_x$.
We fit the measured polarization dependence of the photocurrent
by Eqs.~(\ref{K1}) and~(\ref{K239}) using, besides the ordinate
scaling parameter $\gamma$, the ratio $\eta_{z}/\eta_{\parallel}$
as a fitting parameter. Figure~\ref{fig4} shows that the
MPGE data can be fitted
well by  Eq.~(\ref{K1}) for $\eta_{z}/\eta_{\parallel}\approx 50$
supporting the assumption that the polarization dependence of the
photocurrent can solely be described by the polarization
dependence of absorption.

As demonstrated above, in our experiments the current is caused by
direct inter-subband transitions which  are usually supposed to be
excited by light with the
polarization vector having a nonzero component normal to the QW
plane only (for review see~\cite{stern}). In our consideration the
corresponding absorbance is $\eta_z$. The inter-subband absorption
of light polarized parallel to the  QW plane
$\eta_{\parallel}$ is generally expected to vanish because these
transitions are forbidden by  the dipole selection rules.
However, these rules are valid in the framework of the simple
one-band model only~\cite{Warburton96,IvchenkoTarasenko2004} and
it has been experimentally demonstrated that they are not
rigorous~\cite{Liu1998}. Our data support this conclusion and show
that the absorbance of light with the polarization vector parallel
to the QW plane can be as large as 2~\% of the absorbance of the
light polarized along the
QW growth direction.

\section{Phenomenological analysis}

The MPGE in zinc-blende-type QWs
is related to the gyrotropic properties of the
structures. The gyrotropic point group symmetry makes no
difference between components of axial and polar vectors, and
hence allows an electric current $j_{\alpha} \propto I B_{\beta}$,
where $I$ is the light intensity inside the sample and $B_{\beta}$
are components of the applied magnetic field. Photocurrents which
require simultaneously gyrotropy and the presence of a magnetic
field are gathered in a class of magneto-optical phenomena denoted
as magneto-gyrotropic photogalvanic
effects~\cite{Ivchenkobook2,GanichevPrettl}.
%

The dependence of the photocurrent direction on the light
polarization and orientation of the magnetic field with respect to
the crystallographic axes may be obtained from  symmetry
considerations which do not require knowledge of the microscopic
origin of the effect. Particularly, the MPGE current in response to linearly
polarized radiation and within the linear regime in the magnetic
field  is given by~\cite{Ivchenkobook2,GanichevPrettl,sturman}
\begin{equation} \label{Ch7phen0}
j_\alpha = \sum_{\beta\gamma\delta}
\phi_{\alpha\beta\gamma\delta}\:B_\beta
\frac{E_\gamma E^\star_\delta +  E_\delta E^\star_\gamma}{2} \:,
\end{equation}
where $\bm{\phi}$ is a fourth-rank pseudotensor symmetric
in the last two indices,
$\phi_{\alpha\beta\gamma\delta}=\phi_{\alpha\beta\delta\gamma}$,
and $E_{\gamma}$ are components of the electric field of the
radiation wave in the structure.

We focus on QWs of the C$_{\rm 2v}$ symmetry, which corresponds to
asymmetrically doped (001)-oriented GaAs QW structures studied in
our experiments. In structures of such a symmetry, components of
the MPGE current for, e.g.,  $\bm{B}
\parallel y$ are described by
\begin{equation}
\label{phen1} j_{x} =  [C_1 (e_x^2 + e_y^2) + C_2 (e_x^2 - e_y^2)
+ C_3\, e_z^2)] B_{y} I
\:,
\end{equation}
\begin{equation}
\label{phen2} j_{y} = C_4 e_{x} e_{y} B_{y} I\:,
\end{equation}
where ($e_x, e_y, e_z$) are components of the unit polarization vector
$\bm{e}$ inside the
medium, $\bm{e}$ is assumed to be real for the linearly polarized
radiation; $I$ is the the light intensity inside the
medium related to the electric field of the incident light $E_0$
by $I = (nc E_0^2/4\pi) [(t_s^2 - t_p^2)\sin^2\alpha+t_p^2]$;
$C_1 \div C_4$ are linearly independent
coefficients that can be non-zero in QWs of the C$_{\rm 2v}$ symmetry,
they are related to components of the
tensor $\phi_{\alpha\beta\gamma\delta}$ by
$C_1 \propto (\phi_{xyxx}+\phi_{xyyy})/2$,
$C_2 \propto (\phi_{xyxx}-\phi_{xyyy})/2$,
$C_3 \propto \phi_{xyzz}$,
$C_4 \propto 2\phi_{yyxy}=2\phi_{yyyx}$.

The fact that at
normal incidence ($e_z=0$, $e_x^2 + e_y^2=1$) we observed only a
polarization-independent photocurrent in the direction perpendicular to the
magnetic field demonstrates that the coefficients $C_2$ and $C_4$
in the present experiments are negligibly small as compared to
$C_1$.
At oblique incidence,  the polarization-dependent current contribution determined
by the coefficient $C_3$ is also detected.
We found from experiment that $C_3 \gg C_1$. Such a behaviour is
similar to the polarization dependence of the QW absorbance $\eta$
where, in accordance with Eq.~(\ref{K1}), the radiation polarized
in the QW plane causes weaker optical transitions than that having
nonzero out-of-plane component of the polarization vector. As has
been shown  $\bm{j}$ can be described solely by the polarization
dependence of $\eta$. In this model the parameters $C_1$ and $C_3$
in the phenomenological expression Eq.~(\ref{phen1}) are given by
$C_1=\gamma\eta_\parallel$ and $C_3=\gamma\eta_z$  .

\begin{figure}[!b]
\centering
\includegraphics[width=9cm]{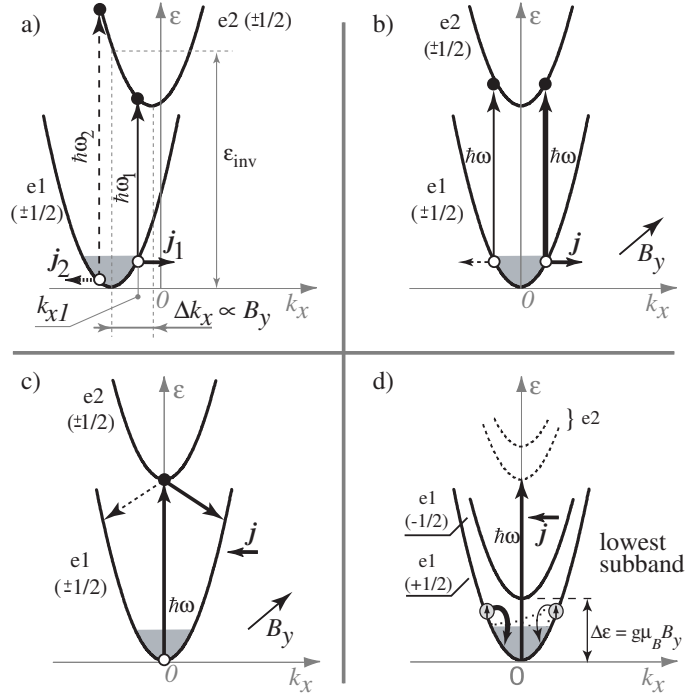}
\caption{Microscopic models of MPGE current formation at
inter-subband transitions. (a) and (b): Current induced by
asymmetric optical excitation due to (a) diamagnetic shift of
subbands in \protect$\bm{k}$-space, (b) linear-in-$\bm{k}$ diamagnetic terms
in the optical transition matrix elements. (c) and (d): Current
induced by asymmetric free carrier relaxation due to (c)
\protect$\bm{k}$-linear diamagnetic terms in the scattering amplitude,
(d) spin-dependent asymmetry of energy relaxation.} \label{fig5}
\end{figure}

\section{Microscopic theory}

Microscopically, the magneto-gyrotropic photocurrents in QW
structures can be of both spin-dependent as well as
spin-independent origin. The proposed spin-dependent mechanisms of
photocurrents include asymmetry of direct optical transitions
between Rashba--Dresselhaus  spin-split branches of the lowest
electron
subband~\cite{Magarill90p2064,Dmitriev91p273,Emelyanov98p810},
spin-dependent asymmetry of the scattering-assisted radiation
absorption by free carriers (Drude absorption)~\cite{naturephysics06}, and
spin-dependent energy relaxation of the electron gas heated by the
radiation~\cite{naturephysics06,PRB07MPGE,IvchenkoAcad1983}.
The spin-independent mechanisms
reported so far
comprise magnetic field induced photocurrents
caused by  a diamagnetic shift of energy
bands~\cite{gorbats,moscow,kucher,Kibis1997,stern}
and diamagnetic  corrections to electron-phonon interaction~\cite{Kibis1998,Kibis2000}.
In this Section we consider step by step various microscopic
mechanisms which can contribute to the photocurrent induced by
optical transitions between the quantum subbands $e1$ and $e2$ in
the presence of an in-plane magnetic field. We assume for
simplicity that the photocurrent is excited at oblique incidence
of  $p$-polarized radiation. This particular case can be
described in a simple single-band model where only the $e_z$
component of the polarization vector induces inter-subband optical
transitions. In the phenomenological theory this contribution  to
the current corresponds to the last term on the right hand side of
Eq.~(\ref{phen1}) proportional to the parameter $C_3$.

\subsection{Diamagnetic mechanisms of the MPGE caused by inter-subband
optical transitions in quantum wells}

Diamagnetic mechanisms of the MPGE   are related to the influence
of the in-plane magnetic field on the orbital motion of carriers
rather than on the electron spin. Microscopically, these
mechanisms originate from the Lorentz
force, which pushes electrons to the right or the left interface
depending on their direction of velocity and, therefore, changes
the electron wave function and energy. Since the Lorentz force is
proportional to the magnetic field and the electron velocity, the
resulting small diamagnetic corrections to the wave function and
energy are linear in $\bm{k}$ as well as linear in $\bm{B}$.

Below, again for simplicity, we consider only linear-in-$\bm{k}$
corrections caused by structure inversion asymmetry. The
corresponding contribution to the effective Hamiltonian induced by
the in-plane magnetic field has the form~\cite{stern}
\begin{equation}\label{H_dia}
H_{dia} = \frac{\hbar e}{m^* c} (B_x k_y  - B_y k_x) z \:,
\end{equation}
where $e$ is the electron charge, $m^*$ is the effective electron
mass, $B_x$ and $B_y$ are components of the magnetic field, and
$z$ is the coordinate operator. The Hamiltonian Eq.~(\ref{H_dia})
corresponds to the vector potential $\bm{A} = (B_y z, -B_x z, 0)$,
$\bm{B} = {\rm rot}\bm{A}$ .

\subsubsection{Current due to $\bm{k}$-linear diamagnetic shift of  subbands}
\label{m1}

It is a well known fact that an in-plane magnetic field
applied to an asymmetric 2D electron gas induces a
spin-independent diamagnetic shift of the electron spectrum in
$\bm{k}$-space in each size-quantized
subband~\cite{stern}. The corresponding corrections to the
electron energies are determined by the diagonal matrix elements
of the Hamiltonian~(\ref{H_dia}) and have the form
\begin{equation}
\delta\varepsilon_{\nu} = \frac{\hbar e}{m^* c} (B_x k_y - B_y
k_x) z_{\nu\nu} \:,
\end{equation}
where $\nu$ is the subband index, $z_{\nu\nu}=\int
\varphi_{\nu}^2(z) z dz$ is the coordinate matrix element,
$\varphi_{\nu}$ is the function of size quantization in the
subband $\nu$ in zero magnetic field. This situation is sketched
in Fig.~\ref{fig5}(a) for $e1$ and $e2$  subbands. We note that,
in spite of the fact that in gyrotropic QWs the spin degeneracy is
removed even in the absence of an external magnetic field, we
neglect both the zero-field spin-splitting and the Zeeman
spin-splitting in the description of diamagnetic mechanisms of the
current formation. The value of the diamagnetic
 shift depends on $z_{\nu\nu}$ and is
generally different for $e1$ and $e2$ subbands. Due to the
relative shift  of the two subbands $\Delta k_x \propto B_y$, the
inter-subband optical transitions induced by the monochromatic
radiation of the photon energy $\hbar \omega_1$ occur only at a
fixed wave vector $k_{x1}$ where the energy of the incident light
matches the transition energy as it is indicated by the solid
vertical arrow in Fig.~\ref{fig5}(a).
Therefore, optical transitions generate  an imbalance of
momentum distribution in both subbands yielding an electric
current. However, a non-equilibrium distribution of carriers in
the upper subband rapidly relaxes due to the very effective
relaxation channel of $LO$-phonon emission, because the energy
of the intersubband optical transition is well above the energy of
$LO$-phonons in $n$-GaAs QWs ($\hbar\Omega_{LO}\approx$ 35~meV).
Thus, the contribution of the $e2$ subband to the electric
current vanishes and the electron flow is determined by
the momentum distribution of carriers in the lowest subband. As
directly follows from this model picture, the variation of the
 incident light frequency causes the
inversion of the current direction. At small photon energy,
$\hbar\omega_1 < \varepsilon_{inv}$, excitation occurs at $k_{x1}$
shifted to the right from the $e1$ minimum resulting in a current
$j_x$ shown by the solid arrow in Fig.~\ref{fig5}(a). The increase
of the photon energy shifts the transition toward negative $k_x$
($\hbar\omega_2 > \varepsilon_{inv}$, dashed vertical arrow in
Fig.~\ref{fig5}(a)) and reverses the direction of the current
shown by the dashed horizontal arrow. The inversion of the current
sign occurs at the photon energy $\hbar \omega =
\varepsilon_{inv}$ corresponding to the optical transitions at the
minimum of $e1$.

Calculations show that, for the case of optical transitions
between the subbands $e1$ and $e2$ and the magnetic field aligned
along $y$, this contribution to the photocurrent has the form
\begin{equation}\label{j_spectrum}
j_x =  (z_{11}-z_{22}) \frac{e^2 B_y}{m^* c}
 \left[ \tau_p^{(2)}\: \eta(\hbar \omega) + (\tau_p^{(1)} -
\tau_p^{(2)})\: \bar{\varepsilon}\: \frac{d \:\eta(\hbar
\omega)}{d\: \hbar \omega} \right] \frac{I }{\varepsilon_{21}} \:,
\end{equation}
where $\tau_p^{(1)}$ and $\tau_p^{(2)}$ are the momentum
scattering times in the subbands, $\eta(\hbar \omega)$ is the QW
absorbance, which is calculated neglecting $\bm{k}$-linear terms
but taking into account the inhomogeneous  spectral broadening of
the inter-subband resonance, $\varepsilon_{21}$ is the
energy separation between the subbands $e1$ and $e2$,  and $\bar{\varepsilon}$ is the
average electron kinetic energy. The latter equals to $k_B T$ and
$E_F/2$ for a non-degenerate and degenerate two-dimensional electron gas,
 respectively, where $T$ is
the temperature and $E_F$ is the Fermi energy.

In accordance to general symmetry arguments, the MPGE current
given by Eq.~(\ref{j_spectrum}) is related to the structure
inversion asymmetry of the QW and vanishes in symmetric structures
where $z_{11}=z_{22}$.
From  Eq.~(\ref{j_spectrum}) follows that, for
the relaxation time of carriers in the excited subband  much
shorter than that in the ground subband, $\tau_p^{(2)} \ll
\tau_p^{(1)}$, the current contribution due to the diamagnetic
shift of the electron subbands is proportional to the spectral
derivative of the QW absorbance. This is similar to the spectral
behaviour of the linear~\cite{Luryi87p2263,Wieck90p463} and
circular~\cite{Shalygin06} photon drag effects as well as the
circular photogalvanic~\cite{PRB03sge,PRBinv} effect caused by
inter-subband transitions in $n$-doped QW structures.

\subsubsection{Current due to $\bm{k}$-linear diamagnetic terms in the
matrix element of optical transitions}
\label{m2}

In the previous Section we considered the photocurrent caused by
linear-in-$\bm{k}$ and linear-in-$\bm{B}$ terms in the energy
spectrum yielding the diamagnetic shift of subbands. Another
diamagnetic contribution to the MPGE, which does not rely on a
band shift, comes from the diamagnetic terms in the matrix element
of inter-subband optical transitions $M$. These terms are also
linear in $\bm{k}$ and $\bm{B}$ and result in  different
probabilities of optical transitions for positive and negative
$k_x$. Indeed, the in-plane magnetic field intermixes the electron
states from the different subbands that leads to asymmetric terms
in the matrix element of optical transitions. This is depicted in
Fig.~\ref{fig5}(b) by vertical arrows of different thickness. This
process leads to an asymmetric distribution of carriers in
$\bm{k}$-space, i.e. to an electrical current. Again the
contribution of the upper subband to the current can be ignored
due to rapid relaxation by emission of optical phonons.

In   first order perturbation theory,  the
wave functions of size quantization  have the form
\begin{equation}\label{functions}
\varphi_{\nu,\bm{k}}(z) = \varphi_{\nu}(z) + (B_x k_y - B_y k_x)
\frac{\hbar e}{m^* c} \sum_{\nu'\neq\nu}
\frac{z_{\nu'\nu}}{\varepsilon_{\nu\nu'}} \varphi_{\nu'}(z) \:,
\end{equation}
where $\varepsilon_{\nu\nu'} =
\varepsilon_{\nu}-\varepsilon_{\nu'}$ is the energy separation
between the subbands. We note that the functions given by
Eq.~(\ref{functions}) with different $\nu$ remain orthogonal at
any wave vector $\bm{k}$ and, therefore, the selection rules for
inter-subband optical transitions are preserved. However, the
in-plane magnetic field adds $\bm k$-linear terms to the matrix
element of optical transitions between the subbands $e1$ and $e2$.
This matrix element can be written as
\begin{equation}\label{M_dia}
M = M_0 \left[ 1 + (B_x k_y - B_y k_x) \frac{\hbar e}{m^* c}
\sum_{\nu\neq1,2} \left(
\frac{z_{\nu1}}{\varepsilon_{\nu1}}\frac{p_{\nu2}}{p_{21}}
-\frac{z_{\nu2}}{\varepsilon_{\nu2}}\frac{p_{\nu1}}{p_{21}}
\right) \right] \:,
\end{equation}
where $M_0 \propto p_{21} e_z$ is the matrix element of the
transitions in zero field, $p_{\nu\nu'}=\int \varphi_{\nu}(z) p_z
\varphi_{\nu'}(z)dz$ is the matrix element of the momentum
operator.

The $\bm{k}$-linear terms in the matrix element of the optical
transitions~(\ref{M_dia}) give rise to an imbalance in the carrier
distribution between positive and negative wave vectors in both
$e1$ and $e2$ subbands, leading to a net electric current.
Calculation shows that the corresponding electric current for the
magnetic field $\bm{B}$ aligned along the $y$ axis is given by
\begin{equation}\label{j_opt}
j_x =  2 (\tau_p^{(1)} - \tau_p^{(2)}) \, \bar{\varepsilon} \,
\frac{e^2 B_y}{m^* c} \sum_{\nu\neq1,2} \left(
\frac{z_{\nu1}}{\varepsilon_{\nu1}}\frac{p_{\nu2}}{p_{21}}
-\frac{z_{\nu2}}{\varepsilon_{\nu2}}\frac{p_{\nu1}}{p_{21}}
\right) \frac{I \eta(\hbar \omega) }{\varepsilon_{21}} \:.
\end{equation}
The photocurrent given by Eq.~(\ref{j_opt}) is present only in
asymmetric QWs. In symmetric QWs, where the wave functions
$\varphi_{\nu}(z)$ are either
even or odd, the current vanishes because the products
$z_{\nu1}p_{\nu2}$ and $z_{\nu2}p_{\nu1}$ are zero for any $\nu$.
In contrast to the mechanism considered in Section~\ref{m1}, the
spectral behaviour of the photocurrent caused by $\bm{k}$-linear
terms in the optical matrix element follows the QW absorbance.
However, estimations show that the contribution described by
Eq.~(\ref{j_opt}) is less than the photocurrent due to the
diamagnetic shift of the subbands given by Eq.~(\ref{j_spectrum})
and, thus, can not be responsible for the observed MPGE.

\subsubsection{Current due to $\bm{k}$-linear diamagnetic terms in the scattering amplitude}
\label{m3}

Besides the asymmetry of optical transitions considered above, an
asymmetry of the subsequent relaxation processes of photocarriers
can also cause an electric current. The relaxation processes
following the intersubband optical excitation include two steps:
electron scattering back from the excited $e2$ to the lower $e1$
subband and subsequent relaxation to the equilibrium within the
$e1$ subband. The former step, electron scattering to the lower
subband, is depicted in Fig.~\ref{fig5}(c) by tilted downward
arrows.  In gyrotropic quantum wells subjected to an external
magnetic field, the matrix element of inter-subband scattering by
static defects or phonons contains an additional term proportional
to $(B_x k_y - B_y k_x)$. Therefore, the scattering rates to final
states with positive and negative wave vectors $k_x$ become
different as reflected in Fig.~\ref{fig5}(c) by arrows of
different thickness. Such an imbalance caused by asymmetry of the
scattering results in an electric current in the ground subband
which is proportional to the applied magnetic field.

To estimate this current contribution we neglect
diamagnetic shifts of the subbands and consider electron
scattering from the state in the second subband bottom
($e2, \,0$) to the state in ground subband ($e1, \,
\bm{k}$) assisted by the most efficient process here of an optical
phonon emission. Taking into account the form of the electron wave
functions in the subbands [see
Eq.~(\ref{functions})], one can derive the matrix
element of inter-subband scattering. For the Fr\"{o}lich mechanism
of electron-phonon interaction~\cite{Ivchenkobook2}, it has the
form
\begin{equation}\label{V_sc}
V_{e1, \bm{k};\,e2, 0}^{(+)} = \frac{C}{q} \left[ Q_{12}(q_z) +
(B_x k_y - B_y k_x) \frac{\hbar e}{m^* c}
\sum_{\nu\neq1}\frac{z_{\nu 1}}{\varepsilon_{1\nu}}\, Q_{2
\nu}(q_z) \right] \delta_{\bm{k},-\bm{q}_{\parallel}} \:,
\end{equation}
where $C$ is a parameter depending on the material,
$Q_{\nu\nu'}(q_z)=\int \varphi_{\nu}(z) \varphi_{\nu'}(z)
\exp(-iq_z z) dz$,
$\bm{q}=(\bm{q}_{\parallel},q_z)$
is the wave vector of the phonon involved,
$\bm{q}_{\parallel}=-\bm{k}$, $k=|\bm{k}|$, and $k =
\sqrt{2m^*(\varepsilon_{21}-\hbar \Omega_{LO})/\hbar}$ due to the
energy and the in-plane quasi-momentum conservation.

The matrix element of inter-subband scattering~(\ref{V_sc})
contains $\bm{k}$-linear terms which lead to asymmetric
distribution of carriers in the ground subband. Calculation shows
that the corresponding electric current for the magnetic field
$\bm{B}$ aligned along the $y$ axis has the form
\begin{equation}\label{j_sc}
j_x =  2 \tau_p^{(1)} \frac{e^2 B_y}{m^* c} \left(
1-\frac{\hbar\Omega}{\varepsilon_{21}} \right)
\sum_{\nu\neq1}\frac{z_{\nu1} \zeta_{\nu}}{\varepsilon_{\nu1}} I
\eta (\hbar \omega) \:,
\end{equation}
where

\begin{equation} \zeta_{\nu} = \frac{\int\int \varphi_1(z)
\varphi_2(z) \varphi_2(z')\varphi_{\nu}(z') \exp{(-|z-z'|k}) \,
dz\,dz'} {\int\int \varphi_1(z)\varphi_2(z)
\varphi_1(z')\varphi_2(z') \exp{(-|z-z'|k}) \,dz\,dz'} \:.
\end{equation}
We note that in this mechanism values of $k$ can be large compared
to that in the mechanism of Section~\ref{m2}.
Similarly to the previous mechanisms of the MPGE, the current due
to asymmetry of scattering~(\ref{j_sc}) is present only in
asymmetric QWs where the products $z_{\nu1}\zeta_{\nu}$ do not
vanish. We also pointed out that processes of energy
relaxation of electrons within the ground subband, which follow
the intersubband scattering, are also asymmetrical in the presence
of in-plane magnetic field ~\cite{Kibis1998,Kibis2000} and
contribute to the current.

The spectral dependence of relaxation currents repeats that of the QW absorbance
$\eta(\hbar\omega)$ which is reasonable for all mechanisms where
details of optical excitation are
lost. Estimations show that these mechanisms can predominate in real structures and
determine the MPGE behaviour.

\subsection{Spin-related mechanisms}

This group of the MPGE mechanisms is based on  spin-dependent
asymmetry of photoexcitation and/or relaxation in the gyrotropic QWs with equilibrium spin
polarization due to the Zeeman effect.
Two spin-related current contributions considered below follow
the spectral dependence of the QW absorbance $\eta(\hbar\omega)$.

\subsubsection{Current due to spin-dependent asymmetry of
energy relaxation}
\label{spin1}

This mechanism is considered in detail
in~\cite{naturephysics06,PRB07MPGE} for electron gas heated by the
Drude-like absorption of the THz radiation. The mechanism is based
on the process of energy relaxation and, therefore, does not
relate on details of optical excitation, besides the strength of
absorption and processes involved in the energy relaxation.

Figure~\ref{fig5}(d) sketches the basic physics of this mechanism.
In the considered case of the radiation absorption due to direct
inter-subband transitions electrons  exited to the $e2$ subband
first return to the ground subband and then rapidly lose their
energy by the emission of optical phonons. After the emission of
optical phonons the energy of electrons gets smaller than $\hbar
\Omega_{LO}$ and the relaxation continues due to the emission of
acoustic phonons. Due to the Zeeman effect, the lowest subband is
split into two spin branches which are unequally occupied. For
simplicity Fig.~\ref{fig5}(d) sketches the process of energy
relaxation of hot electrons for only the lowest spin-up branch
($s=+1/2$). Energy relaxation processes are shown by curved
arrows. Usually, energy relaxation via scattering of electrons is
considered to be spin-independent. In gyrotropic media, however,
spin-orbit interaction adds asymmetric spin-dependent term to the
scattering probability. These terms in the scattering matrix
element are proportional  to components of
$[\bm{\sigma}\times(\bm{k}+\bm{k}^\prime)]$, where $\bm{\sigma}$
is the vector composed of the Pauli matrices, $\bm{k}$ and
$\bm{k}^\prime$ are the initial and scattered electron wave
vectors in the ground subband (we consider here only the
spin-dependent contribution induced by heteropotential asymmetry).
%
Due to spin-dependent scattering, transitions to positive and
negative $k_x^\prime$-states occur with different probabilities.
Therefore hot electrons with opposite $k_x$ within one branch have
different relaxation rates. In Fig.~\ref{fig5}(d) this difference
is indicated by arrows of different thickness. 
This in turn yields a net electron flow, $\bm{i}_{\pm 1/2}$,
within each spin branch. Since the
asymmetric part of the scattering amplitude depends on spin
orientation, the probabilities for scattering to positive or
negative $k^\prime_x$-states are inverted for spin-down and
spin-up branches.
Without magnetic field, the charge currents $\bm{j}_+ =
e\bm{i}_{+1/2}$ and $\bm{j}_- = e\bm{i}_{-1/2}$ have opposite
directions because $\bm{i}_{+1/2} = -\bm{i}_{-1/2}$ and
 cancel each other.
%
However, the external magnetic field changes the relative
population of spin branches and,
therefore, lifts the balance between $\bm{j}_+$  and  $\bm{j}_-$
yielding a net electric current.
The photocurrent magnitude can be estimated as
\begin{equation}
\label{spineq}
j \sim e \tau_p^{(1)} S^{(0)} \, \frac{\xi}{\hbar}  \, I
\eta(\hbar\omega) \:,
\end{equation}
where $S^{(0)}=g \mu_B B /(4k_B T)$ is the equilibrium spin
polarization for the Boltzmann distribution, and $\xi$ is
a parameter standing for the ratio of spin-dependent to
spin-independent parts of electron-phonon interaction.

\subsubsection{Current due to asymmetry of  spin relaxation
(spin-galvanic effect)}
\label{spin2}

For completeness we also give an estimation for the second
possible spin-dependent mechanism, previously considered
by us in Ref.~\cite{Belkov2005}. It is based on the asymmetry of
spin-flip relaxation processes and represents in fact the
spin-galvanic effect~\cite{Nature02} where the current is linked
to non-equilibrium spin polarization
\begin{equation}
j_i = Q_{ii'} (S_{i'} - S^{(0)}_{i'})\:. \label{Q}
\end{equation}
Here ${\bm S}$ is the average non-equilibrium electron spin and
${\bm S}^{(0)}$ is its equilibrium value, see Eq.~(\ref{spineq}).
In contrast to the mechanisms considered in
Sections~\ref{m1}-\ref{m3} this mechanism requires spin-flip
processes
%
%
together with a \textit{non-equilibrium} spin polarization. A
\textit{non-equilibrium} spin polarization results from the
photoinduced depolarization of electron spins in the system with
equilibrium polarization caused by the Zeeman effect. Indeed, due
to the fact that in equilibrium electrons preferably occupy the
lower spin branch,
optical transitions, being proportional to the electron
concentration, predominantly excite this branch. These optically excited electrons under energy
relaxation return to both spin branches
resulting in non-equilibrium population of the
branches. The following spin relaxation  results in the
spin-galvanic current (see Ref.~\cite{Nature02}).

An estimation of this photocurrent for the D'yakonov-Perel spin relaxation mechanism
yields
\begin{equation}
j \sim e \tau_p^{(1)} \,S^{(0)} \, \frac{\beta^{(1)}}{\hbar}
\frac{I \eta(\hbar\omega)}{\varepsilon_{21}} \:,
\end{equation}
where $\beta$ is the constant of $\bm{k}$-linear spin-orbit
splitting of the subband $e1$.

\section{Conclusions}

Summarizing our experiments and the microscopic theory developed
here we emphasize that the resonance spectral behaviour of the
observed magnetic field induced photocurrent following the
inter-subband absorption profile  proves that the photocurrent is
due to direct inter-subband transitions. As an interesting
feature, the resonant photocurrent is also detected at
normal incidence of the radiation on (001)-grown QWs. This
result demonstrates that the selection rules conventionally
applied for inter-subband optical transitions  are not rigorous.
Considering that the photocurrent is proportional to the
QW absorbance, we obtain from the experiment that the ratio of
absorbance for light polarized along the growth direction  $z$
and in the QW plane is $\eta_{z}/\eta_{\parallel} \approx 50$.
%
Surprisingly, the MPGE photocurrent is not caused by the
diamagnetic shift of subbands (see Section~\ref{m1}). Such a
current has been previously reported  for inter-band absorption
and is expected to be strong compared to spin-dependent mechanism
because of its non-relativistic nature. However, the absence of
spectral inversion of the photocurrent at intersubband absorption resonance
 unambiguously rules out this mechanism. Three
other possible mechanisms of current formation developed here
cannot be distinguished qualitatively. Based on quantitative
estimations of the mechanisms given in Sections from~\ref{m2} to
\ref{spin2} we attribute the MPGE current to $\bm
k$-linear diamagnetic terms in the scattering
matrix element yielding asymmetric relaxation of carriers in $\bm
k$-space (Section~\ref{m3}). Finally we note that diamagnetic and
spin-dependent mechanisms might be qualitatively distinguished in
structures where the $g$-factor can be varied like in dilute
semimagnetic materials. Indeed, while spin-dependent
mechanisms are proportional to the $g$-factor, diamagnetic
mechanisms are independent of the Zeeman splitting.

\section*{Acknowledgements}

 We thank E.L. Ivchenko for helpful discussions. This work was supported by the Deutsche
For\-schungs\-ge\-mein\-schaft
through GA 501/6, SFB 689, SPP 1285,
the RFBR, programs of the RAS and Russian Ministry of Education and Science,
Russian President Grant for young scientists, and the Russian Science Support Foundation.




\end{document}